\documentclass[[12pt]{article}
\usepackage{verbatim}
\usepackage[english]{babel}
\usepackage{pifont}
\usepackage{graphicx}
\usepackage{natbib}
 \usepackage{mathptmx}      % use Times fonts if available on your TeX system
\usepackage{latexsym}

\begin{document}

% Include your paper's title here

\title{Can we predict long-run economic growth?}

% Place the author information here.  Please hand-code the contact
% information and notecalls; do *not* use \footnote commands.  Let the
% author contact information appear immediately below the author namesf
% as shown.  We would also prefer that you don't change the type-size
% settings shown here.

\author
{Timothy J. Garrett }

%\institute{Department of Atmospheric Sciences\\
%University of Utah\\
%Salt Lake City, USA\\
%\email{tim.garrett@utah.edu.}
%}

% Include the date command, but leave its argument blank.

\date{}

%%%%%%%%%%%%%%%%% END OF PREAMBLE %%%%%%%%%%%%%%%%

\maketitle
\begin{abstract}
For those concerned with the long-term value of
their accounts, it can be a challenge to plan in the
present for inflation-adjusted economic growth
over coming decades. Here, I argue that there
exists an economic constant that carries through
time, and that this can help us to anticipate the
more distant future: global economic wealth has a
fixed link to civilization's overall rate of energy consumption from all sources;
the ratio of these two quantities has
not changed over the past 40 years that statistics
are available. Power production and wealth rise equally
quickly because civilization, like any other
system in the universe, must consume and
dissipate its energy reserves in order to sustain its
current size. One perspective might be that
financial wealth must ultimately collapse as we
deplete our energy reserves. However, we can also
expect that highly aggregated quantities like
global wealth have inertia, and that growth rates
must persist. Exceptionally rapid innovation in the
two decades following 1950 allowed for
unprecedented acceleration of inflation-adjusted
rates of return. But today, real innovation rates are
more stagnant. This means that, over the coming
decade or so, global GDP and wealth should rise
fairly steadily at an inflation-adjusted rate of about
2.2\% per year.   
\end{abstract}

\section{Introduction}

Our financial accounts seem to change
unpredictably according to the actions of
individuals, organizations and governments.
Because the range of human behavior can be so
diverse and out of our control, it seems that there
is an exceptionally broad range of future societal
outcomes. Anticipating long-term economic
conditions anything more than a year away seems
daunting at best.

Forecasting future human behavior becomes
relevant where the goal is to provide society with
forecasts of climate change. Through the
combustion of fossil fuels, our economic
activities have been slowly increasing
atmospheric greenhouse gas concentrations.
Consequent changes in climate patterns remain
modest. But, perhaps several decades from now,
global warming will become an important drag on
economic growth \citep{IPCC_WG22007}. 

For those concerned with their personal financial accounts, the issue is more
about anticipating inflation-adjusted rates of
return, sometimes as much as a
decade or more ahead. While, diversification can
help financial portfolios to weather short-term
fluctuations in market valuations, the optimal
strategy for the longer term is less clear. For
example, regardless of investment strategy, Baby
Boomers have generally prospered from a rising
economic tide that has lifted most boats over the
past half-century. Yet many worry that such
extraordinary overall gains cannot persist
indefinitely. All tides subside. Broad economic
gains can be lost.

Are we confined to hoping for the best but preparing for the worst? Or can we at least plan
ahead for the future by making constrained
predictions for where our net worth and rates of energy consumption are headed? In
a recent paper focused on the implications for climate change, I proposed that we can, provided
that we are willing to take a broader view by
considering slow changes in the economy as a
global whole \citep{GarrettCO2_2009}. 

\section{A link between economics and physics}
\begin{figure}
\includegraphics[width=11cm]{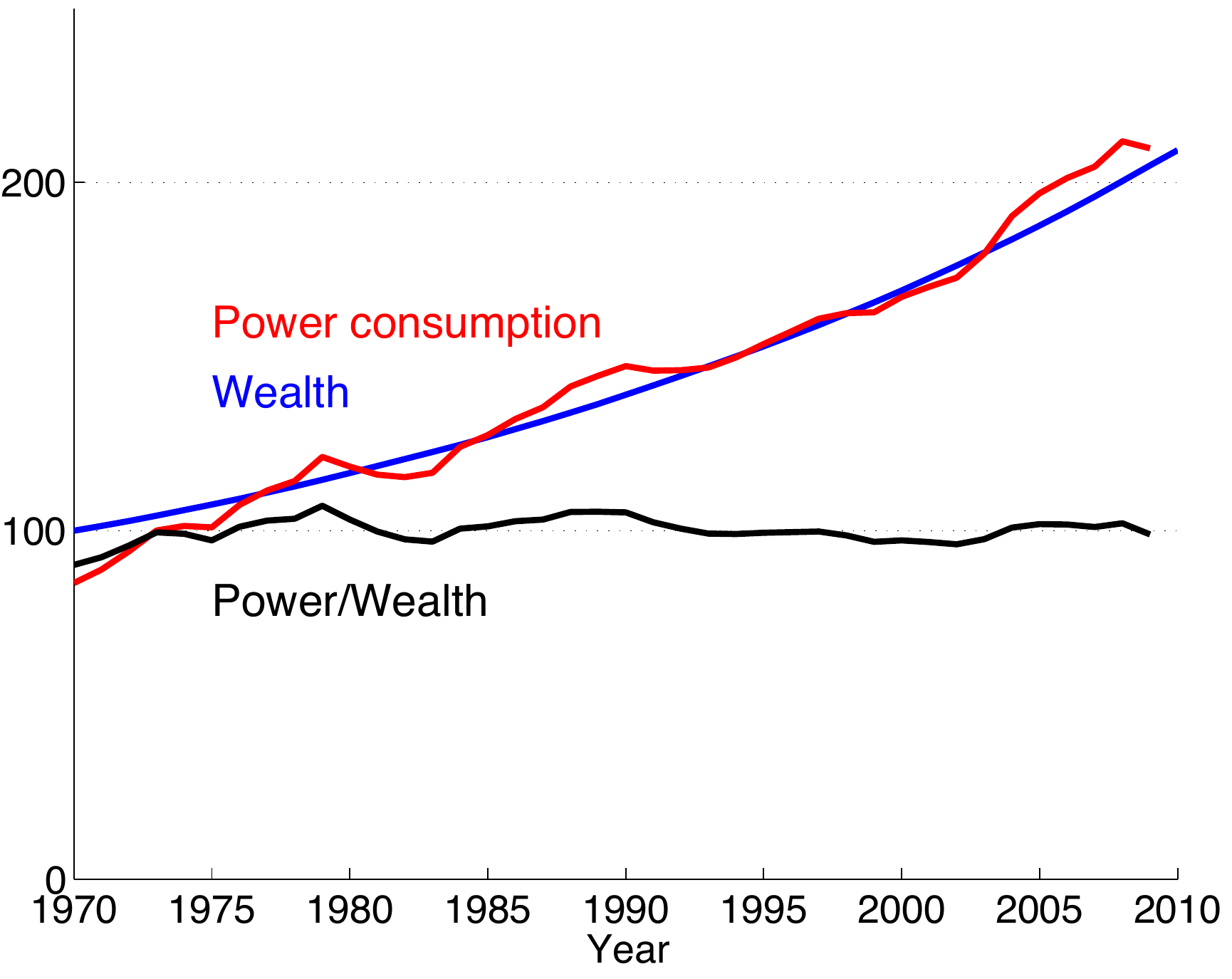}

\caption{\label{fig:growth}Since 1970, global wealth as defined by Eq. \ref{eq:wealth} (blue), global power production (red), and the ratio of power to wealth (black). Wealth is referenced to 100 in 1970. Power production is derived from all possible sources, including fossil, nuclear, biomass, and renewables.}

\end{figure}

Physical reasoning suggests that a very broad measure of civilizationÕs total fiscal wealth should have a
fixed link to its overall rate of primary energy
consumption, or power production, independent of time. Observations
seem to support this hypothesis to a remarkable
degree (Figure \ref{fig:growth}). The
implication is that global wealth will continue to
rise for as long as power production can
continue to grow. Otherwise, if resources ever
become so constrained that the capacity to consume energy falls,
global wealth must enter a phase of collapse. 

Before elaborating further on this new economic growth model, it is worth examining traditional macro-economic models that focus on human labor and creativity as the motive economic forces. Almost all economists treat
``{}human'' capital, or labor, and ``{}physical'' capital as two totally distinct quantities. Labor and
capital combine in a complex way to enable
economic production. A small part of economic
production is a savings that can be carried into the
future as added physical capital. But most
production is siphoned away by people through
their consumption of such things as food and
entertainment. Once something is consumed, it
has no potential to influence future economic
activities.

While this model is certainly logical, from the
standpoint of physics, it seems strange because it
appears to both ignore and violate the most
universal of laws: the Second Law of
Thermodynamics. The Second Law is familiar to
many for its statements about entropy production.
Perhaps the best known is that the universe, taken
as a whole, inescapably slides towards increasing
disorder. 

But the Second Law also demands that nothing can do anything without consuming concentrated energy, or fuel, and then dissipating it as unusable waste heat. For example, the Earth ``{}consumes'' concentrated sunlight to power weather and the water cycle, and then radiates diffuse, unusable thermal energy to the cold of space. 

Like the weather in our atmosphere, all economic
actions and motions, even our thoughts, must also
be propelled by a progression from concentrated
fuel to useless waste heat. The economy would
grind to a halt absent continued energetic input.
Machines slow down; buildings crumble; people die; technology
becomes obsolete; we forget. Civilization must
constantly consume in order to sustain itself
against this constant loss of energy and matter.

Even we as individuals consume the energy in food at an average rate of about 100 Watts. This sustains and builds the joint activities of our brain, heart, lungs, and other body functions. We must keep eating to regenerate dead cells and offset the constant loss of heat through our skin. 

Taken as a whole, civilization is no different, except that
after centuries of growth, it is rather large and
wealthy. Today, sustaining all of our activities
requires continuous consumption and dissipation
of about 17 trillion Watts of power, or the
equivalent of 17,000 one Gigawatt power plants.
We burn fossil fuels, split uranium nuclei and tap
the potential energy in rivers, sunlight and wind.
About 4\% of this energy is dissipated by our 7 billion
bodies. The rest powers our agriculture, buildings
and machines. Once consumed, all energy is
ultimately dissipated as waste heat. If energy
consumption ever ceased, our machines would
stop, and we would all die. Certainly, economic
wealth would be zero.

\section{Treating the economy as a global whole}
The idea that the economy is sustained by power production has certainly been discussed by others \citep{AyresWarr2009}. However, there is a particularly
beautiful corollary of the Second Law whose
implications for the economy have largely been missed in these treatments. This is the
statement that nothing can be isolated: all of space
and time are linked. Nothing happens
spontaneously, and all actions from the past have
some influence on the present and future. Equally,
no sub-component of the universe can be
completely separated from interactions with any
part of the rest. However, remote or slow the
interactions may be, all parts are connected to and
interact with all others.

What this means for how we model societal behavior is perhaps best expressed by the Elizabethan poet John Donne, ``{}No man is an island, entire of itself. Each is a piece of the continent, a part of the main.''  Through international communications and trade,
ourselves, our ideas, education and relationships
all form a vibrantly interacting and changing
whole that is completely integrated with our
transportation routes, communication networks,
factories, buildings and databases.

In other words, all elements of civilization work
together. No matter how distant, no element of
economic production can be isolated from any
other. We are all part of a vibrant organism we call
the global economy. 

So where traditional economic models allow a portion of real production
to simply disappear to the past due in the form of ``{}consumption", the Second Law would
suggest that this is impossible. For one, humans
are not isolated, but instead inextricably linked
to the rest of the organism's overall structure. Human consumption is not cleanly separable from energy 
dissipation by anything else. For another, past actions are carried forward to the present and future. 
Civilization's history of consumption has sustained it against
dissipation and decay, and nurtured it
forward such that it is able to continue to consume in the
present. For example, Ancient Greece sustained an
architectural tradition that has been carried
forward to current designs. Or, entertainment
consumed a hundred years ago sustains a cultural
tradition that influences our choices today. Our ancestors could not have procreated to enable
our own existence if they had not consumed and dissipated the energy in food.

\section{A new model for economic growth}
In an economic growth model, the above
arguments are simply expressed by a hypothesis
that ``{}Wealth is Power''. We are sustained by a
consumption of energy. A portion of inflation-adjusted
economic output cannot disappear. Rather
all of production is returned to wealth, where wealth is defined
very generally as human and physical capital combined.

Unlike traditional economic treatments, real
production can neither be siphoned off to consumption by humans
alone, nor to the past. It has nowhere else to go
but to ``{}produce''. Put another way, wealth is the accumulation of all past production, adjusting for inflation.
The current inflation-adjusted global GDP is like a ``{}rate of return" on global
wealth since, like returns in a bank account, it adds to whatever wealth currently exists. 

The tie to physics is the hypothesis that this accumulated wealth is directly
proportional to civilization's total capacity to consume energy or produce power. And just as 
we gain weight when we eat too much, only when
energy consumption exceeds dissipation can a
convergence of flows allow for civilization
expansion and a positive inflation-adjusted
economic output or GDP (see Appendix for the
mathematical details).

Crucially, this hypothesis is falsifiable. In other
words, it is sufficiently simple, transparent and
easy to test, that it could potentially be discarded
based on observational evidence. I tested this
hypothesis using statistics for world GDP and
energy consumption that are available for each
year from 1970 onward, together with more sparse
estimates for world GDP that extend back to
1 AD \citep{GarrettCO2_2009}. 

What these data show is that, for each year between 1970 and 2009, the ratio of current power production to our historical accumulation of wealth has barely deviated from a constant 7.1 Watts per thousand inflation-adjusted 2005 US dollars (Figure \ref{fig:growth} and Table \ref{tab:Measured-values}). The standard deviation has been only 3\% during a period when global power production and wealth have increased by 120\% and world GDP has risen 230\%. Wealth and power production are not merely correlated, any more than photon frequency and energy are correlated in the formula $E={h}\nu$. Rather, like frequency and energy, the ratio of the two appears to be fixed.

It seems extraordinary, but the implication is that
we can begin to think of seemingly complex
human systems as simple physical systems. Our
collective fiscal wealth is an alternative and very
human measure of our capacity to power our
society through the consumption of fuel. Our
total assets, including ourselves, our relationships
and our knowledge, are inseparable from our
collective capacity to consume our primary
reserves of coal, oil, natural gas, nuclear fuels and
renewables. Both will rise and fall together.
 
\begin{table}[htp]
\caption{\emph{\small \label{tab:Measured-values}For select years since 1970, measured values for the global
power production (trillion Watts), global real wealth (trillion
2005 MER USD)}{\small ,}\emph{\small{} the ratio of power production to wealth (Watts
per thousand 2005 MER USD), global real GDP (trillion 2005 MER USD per
year) and the rate of return on wealth defined by GDP/Wealth  (\% per year). }}

\begin{tabular}{cccccccccc}
\hline 
 & \textsf{\small 1970} & \textsf{\small 1975} & \textsf{\small 1980} & \textsf{\small 1985} & \textsf{\small 1990} & \textsf{\small 1995} & \textsf{\small 2000} & \textsf{\small 2005} & \textsf{\small 2009}\tabularnewline
\hline
{\footnotesize Power production (TW) } & {\footnotesize 7.2} & {\footnotesize 8.3} & {\footnotesize 9.6} & {\footnotesize 10.2} & {\footnotesize 11.6} & {\footnotesize 12.1} & {\footnotesize 13.1} & {\footnotesize 15.2} & {\footnotesize 16.1}\tabularnewline
{\footnotesize Power/Wealth } & {\footnotesize 6.4} & {\footnotesize 6.9} & {\footnotesize 7.3} & {\footnotesize 7.2} & {\footnotesize 7.5} & {\footnotesize 7.1} & {\footnotesize 6.9} & {\footnotesize 7.2} & {\footnotesize 7.0}\tabularnewline
{\footnotesize GDP (Trillion \$/year)} & {\footnotesize 15.3} & {\footnotesize 18.4} & {\footnotesize 22.2} & {\footnotesize 25.3}& {\footnotesize 30.2} & {\footnotesize 33.5} & {\footnotesize 39.7} & {\footnotesize 45.7} & {\footnotesize 49.1} \tabularnewline
{\footnotesize Rate of return (\%/year)} & {\footnotesize 1.37} & {\footnotesize 1.53} & {\footnotesize 1.70} & {\footnotesize 1.78} & {\footnotesize 1.94} & {\footnotesize 1.96} & {\footnotesize 2.10} & {\footnotesize 2.18} & {\footnotesize 2.14}\tabularnewline
\hline
\end{tabular}

\end{table}
.

\section{Precision versus predictability in economic forecasts}

For those concerned with climate change or the
long-term value of their financial accounts, a link between
economics and physics has some important
implications. We
can anticipate inertia in global consumption and
economic growth. Our current consumption and
wealth are inextricably tied to past production, but
the past is unchangeable. Absent some sort of
severe external shock, near-term reductions in
energy consumption and wealth are implausible
because they would somehow require civilization
to {}``forget'' its past.

Assuming that economic consumption and growth
will persist in the near term may seem rather
obvious to some. But what may be less well
recognized is that there are mathematical and
physical constraints to growth. For those who study
the evolution of physical systems, a term that is
often used here is {}``reddening''. This is a convenient
way of expressing that it is the most slowly varying,
low frequency and {}``red'' (rather than blue)
components of past variability in a system that most
strongly influence its present behavior.
 
For example, seasonal temperature trends
normally have a stronger influence on daily
high temperatures than shorter term weather
variability. Or, 20 years of growth through
childhood and adolescence tends to have a greater
influence on our daily food consumption than how
much we ate yesterday. Surprises can happen, of
course. For us, there is always the potential for
accident or a disease. Still, the natural tendency
for growth is for it to be slow and steady.

Equally, the global economy's current capacity to
consume and grow has evolved from thousands of
years of human development, through the creation
of subsequent generations, as well as the
construction of farms, towns, communication
networks and machines. While everything does
slowly decay or die, the past can never be entirely
erased. Even our most distant ancestors have
played a role in our current economic and social
well-being. By now, civilization has enjoyed a
rather lengthy past, and we can count on this
accumulated inertia to carry us into the future. 

Certainly, it is still possible that countries will rise and fall, but globally aggregated economic wealth should continue to enjoy recent inflation-adjusted rates of return. Even in 2009, during the depths of the Great Recession, 2.14\%  was added to total real global wealth (Table \ref{tab:Measured-values}), only slightly down from the historical high of 2.26\% in 2007. And we continue to grow our
power production at similar rates. It is probably a safe bet to assume that similarly high rates of return will persist over the coming decade. 

The main point is that persistence in trends is an effective tool for forecasting, but most especially when applied to highly {}``reddened" variables like global wealth and energy consumption that are aggregated over time and space. When predicting the evolution of any system, there is always a trade-off. It is always easier to make forecasts provided that we are willing to sacrifice temporal and spatial resolution. 

Predicting the weather next week can be almost
impossible. But forecasting northern hemisphere
average temperatures this coming winter is
actually quite easy: history is an excellent guide.
Similarly, it is very difficult to predict a small
companyÕs stock value next week; but
extrapolating trends in globally-aggregated wealth
can plausibly be done for as much as a decade
hence.

\section{Innovation and increasing rates of return}

To reiterate, available statistics show that wealth, when it is integrated over the entire global economy, and integrated over the entire history of economic wealth production, has been related to the current rate of global primary energy consumption through a factor that  has been effectively constant over nearly four decades of civilization growth. The implication is that aggregated civilization wealth and consumption has inertia, and therefore its current growth rate is unlikely to cease in a hurry.

\begin{figure}
\includegraphics[width=11cm]{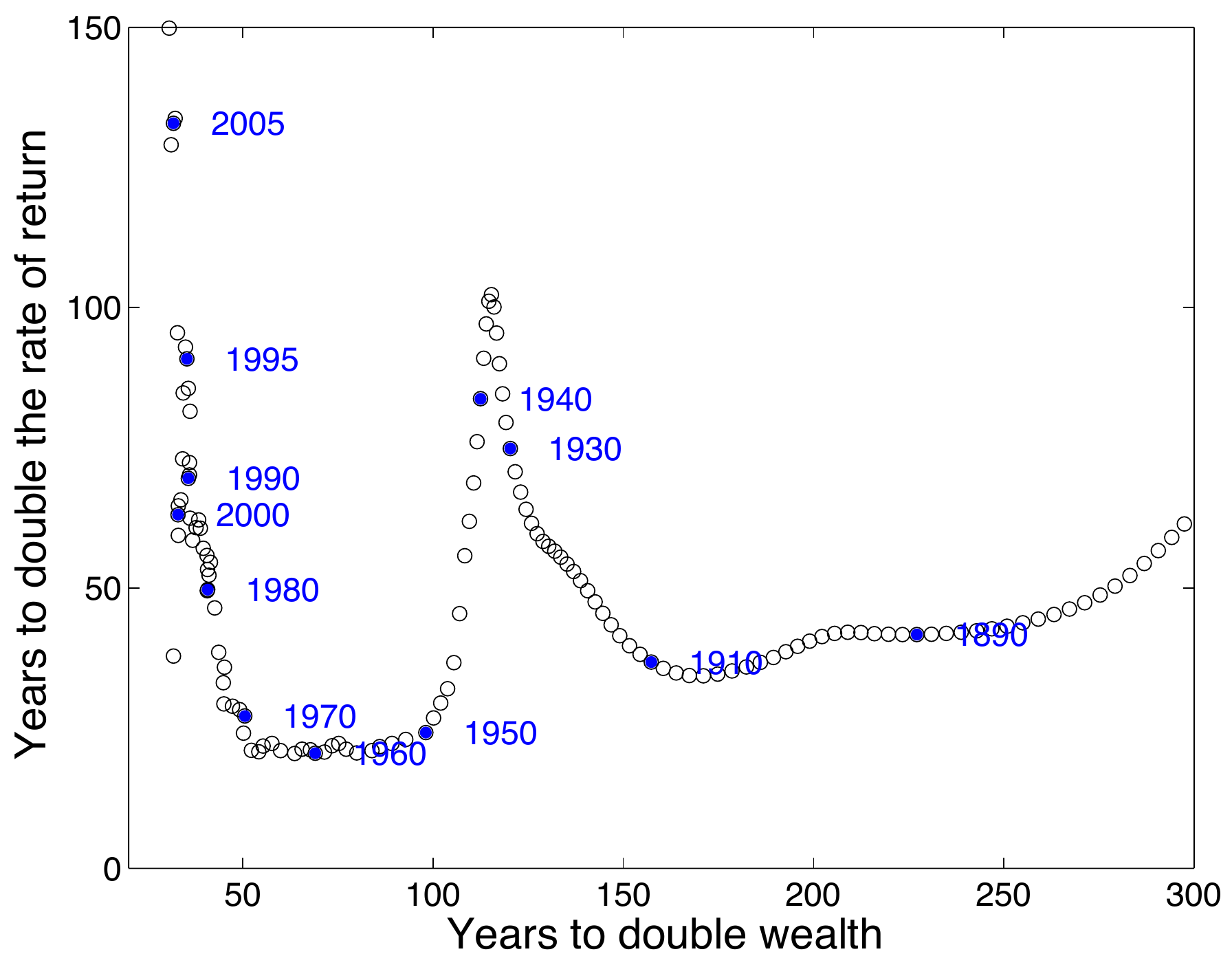}

\caption{\label{fig:doubling}Adjusting for inflation, the time for global wealth's rate of return to double (calculated as a decadal running mean), versus the doubling time for wealth itself. Select years are shown for reference. }

\end{figure}

Yet the global rate of return on wealth does change, even if slowly. Historical statistics shown in Table 1 and Figure \ref{fig:doubling} indicate that, over the past century or so, there has been a long term tendency for wealth to double over ever shorter intervals. In the late 1800s, doubling global wealth would have taken about 200 years based on then--current rates of return. Today this takes just 30 years. As a whole, the world is getting richer faster. 

I use the word innovation to describe this acceleration of inflation-adjusted rates of return because it represents the capacity of civilization as a whole to beat mere inertia. Adjusting for inflation is important here, because it is not always evident that any investment in innovation will pay off. If investing in human creativity does not lead to true innovation, then it is a waste of effort that could have otherwise contributed to previously attained rates of growth. But, real innovations provide a jump in rates of return that civilization can carry forward into the foreseeable future.

Globally, innovation has come in fits and starts. Figure \ref{fig:doubling} shows that innovation has had two golden periods over the past two centuries. The first was during the Gilded Age or ``{}Belle Epoque'' of the late 1800s and early 1900s, when resource expansion and  technological discoveries allowed the rate of return to double in just 40 years. Then again, in the baby boom period between 
1950 and 1970, the rate of return doubled in the remarkably short timespan of just 20 years. %

By contrast, both the 1930s and the past decade have been characterized
by much more gradual inflation-adjusted innovation rates.  Even though wealth is now doubling more quickly
than ever before in history, for the first time since the Great Depression the rate of return is no longer increasing. 

Why has the passage of history been characterized by economic ``{}fronts'' on global scales, with rapid innovation giving ultimately giving way to stagnation? 

Here again, physical principles can provide guidance. Given that inflation-adjusted wealth and energy consumption appear to be linked through a constant, the identical question is asking what enables energy consumption to accelerate. 

Conservation laws from thermodynamics tell us that rates of innovation and growth should be largely controlled by the balance between how fast civilization discovers new energy reserves and how fast it depletes them \citep{Garrettmodes2012}. For example, it is easy to imagine that access to important new coal or oil reserves
in the late 1800s and around 1950 allowed civilization to capitalize on human creativity in ways that were previously impossible. 

Today, we continue to discover new energy reserves, but perhaps not sufficiently quickly. We are now very large and we are depleting our reserves at the most rapid rate yet. Increased competition for resources may be constraining our capacity to turn our creativity and knowledge into real innovation and accelerated global economic growth. 

\section{Conclusions}

I have described here a constant that links a very general representation of the world's total economic wealth to civilization's power production capacity. Because this constant does not change with time, physical principles can be applied  to estimate future global-scale economic growth over the long-term without having to explicitly model the exceptionally complex internal details of people and their lifestyles. 

There have been criticisms of this approach, which have stated that {}``Economic systems are not the same as physical systems, and we shouldn\textquoteright{}t model them as
if they are'' \citep{Scher2010}. Yet, civilization is undeniably part of the physical universe. It is difficult to imagine how we aren't fundamentally constrained by physical laws. 
At the very least, appealing to physics appears to make the job of economic forecasting more transparent, simple and scientifically robust. 

Here, I have made the argument that recent rates of return are most likely to persist in such highly aggregated quantities as the global economy. We can make quite distant estimates of future growth, but only if we are willing to sacrifice resolution. While this does not specifically help us to predict trajectories for specific countries or economic sectors, we might anticipate that slower than average rates of return in one nation or sector should be balanced by faster growth elsewhere.

It must be kept in mind that exponential growth trends cannot continue unabated. Wealth is tied to power production and therefore to resource consumption. Sooner or later, civilization must face up to reserve depletion or environmental degradation. 

But, for as long as these impacts remain manageable, we can anticipate that global economic wealth, GDP, and energy consumption will continue to grow at recently observed rates. The qualification is that rates of return are unlikely to rise as fast as they did in the decades following the 1950s. Rather, for timescales significantly less than the wealth doubling time of 30 years -- perhaps a decade -- the forecasted inflation-adjusted global rate of return should average a fairly steady 2.2\% per year. 

For those thinking even further ahead, inflation-adjusted rates of return should be guided by whether there is net depletion or expansion of our primary energy reserves. It might help to think of our energy reserves as the retirement account for civilization as a whole. Discovering new energy reserves today expands our collective accounts. But having sufficient reserves for the long-term requires that we not ``{}spend down'' what we have discovered too quickly. What we consume today must also be balanced against what we have left to consume in the future.
\subsection*{Acknowledgments}
This work was supported by the Kauffman Foundation, whose views it does not claim to represent.

\newpage

	\begin{center}
   \bf\large{Appendix: A physical model for economic growth}
    \end{center}
The model for long-term economic growth used here is based on a core hypothesis, motivated by the Second Law of Thermodynamics,
that energy consumption is required to power economic activities, and
that all elements of civilization must be considered as part of a larger whole. No portion of production can disappear to the past as human consumption, with no influence on the future \citep{GarrettCO2_2009}.

Expressed mathematically, all economic production, once adjusted for inflation, contributes to global wealth or capital, so that
\begin{equation}
\frac{dC}{dt} = Y \label{eq:dCdt}\end{equation}
or, in integral form:
\begin{equation}
C=\int_{0}^{t}Y\left(t'\right)dt'\label{eq:wealth}\end{equation}
where $C$, the inflation-adjusted
economic value (or civilization wealth) $C$ is calculated from the
time-integral of global economic production (or GDP) $Y$, adjusted
for inflation at 2005 market exchange rates (MER), and aggregated over
the entirety of civilization history. While not described here, under this approach, the GDP deflator is related to capital decay in traditional approaches \citep{Garrettcoupled2011}.

Global economic wealth $C$ is sustained by the instantaneous rate of primary energy consumption by civilization
$a$ through a constant $\lambda$ \begin{equation}
a=\lambda C \label{eq:alambdaC}\end{equation}
Taking $a$ to be in units of Watts (Joules per second), and $Y$ in units of 2005 MER
US dollars per second, available statistics indicate that $\lambda$ is indeed constant, with a measured value
of 7.1 Watts per 1000 dollars. The standard deviation over the past forty years has been just 3\%. 

So, at least over the long-run, sustaining a fiscally measurable inflation-adjusted world GDP requires a \emph{growing} power production. From Equations \ref{eq:dCdt} and \ref{eq:alambdaC}, the production function for the growth model is
\begin{equation}
Y = \frac{1}{\lambda}\frac{da}{dt}\label{eq:GDP}\end{equation}

Also, from Equation \ref{eq:alambdaC}, global wealth $C$ and energy consumption $a$ must rise at the same rate. This can be expressed in a variety of ways, all of which follow from Eqs. \ref{eq:dCdt} to \ref{eq:GDP}:
\begin{equation}
\rm{Rate\,of\,Return} =\eta=\frac{Y}{C}=\frac{1}{C}\frac{dC}{dt}=\frac{1}{a}\frac{da}{dt}=\frac{Y}{\int_{0}^{t}Y\left(t'\right)dt'}\label{eq:eta}\end{equation}
While the rate of return $\eta$ is tied to growth in power production, it is also an expression of the ratio current economic output to the time integral of past economic output, adjusted for inflation.

\subsection*{Productivity and GDP growth}
Very often in economic studies, one sees the ratio $f=Y/a$, which is termed the {}``energy productivity'' or {}``energy efficiency'' of the economy since it represents how well economies turn energy consumption $a$ into economic production $Y$. Since $a = \lambda{C}$ and $\eta = Y/C$, it follows that: 
\begin{equation}
\rm{Rate\,of\,Return} = \eta=\lambda{f}\label{eq:etaf}\end{equation}
So, since $\lambda$ is a constant, the rate of return on wealth is proportional to the economy's energy productivity, and only by increasing energy productivity can the rate of return increase.

When the rate of return increases, this can be interpreted as a consequence of  {}``real innovation'' since, from Eq. \ref{eq:etaf} it corresponds to greater energy productivity $f$. The innovation rate is: 
\begin{equation}
\rm{Innovation\,rate} = \frac{1}{\eta}\frac{d\eta}{dt} = \frac{1}{f}\frac{df}{dt} \label{eq:innovation}\end{equation}
If innovations improve energy productivity $f$, then they accelerate growth and lead to higher rates of return on wealth. Curiously, from Eq. \ref{eq:alambdaC}, they also lead to increasing power production $a$. 

The possibility that energy efficiency gains might ``{}backfire''  has been hotly disputed, and it is contrary to what is normally assumed (see also \cite{Saunders2000,Jevonsbook2007,Owen2010}). Indeed, it does seem quite counter-intuitive that improving energy efficiency leads to more rather than less power production.  But, the system must be considered as a whole. Innovation allows civilization to collectively expand at a more rapid pace into the energy resources that sustain it, and this allows it to produce more power.

Perhaps the most commonly quoted macro-economic statistic is the real GDP growth rate. In the framework here, this can be expressed very simply as the sum of the current rate of return, and the innovation rate or acceleration of the rate of return \begin{equation}
\frac{1}{Y}\frac{dY}{dt}=\rm{rate\,of\,return\,+innovation} = \eta+\frac{d\ln\eta}{dt}\equiv\lambda f+\frac{d\ln f}{dt}\label{eq:dlnPdt}\end{equation} 
In fact, since the current productivity arises from past innovations (i.e, $f = \int_0^t{df/dt' dt'}$), the history of innovation is the sole motivating force driving current GDP growth.

To illustrate, the mean energy productivity $f=Y/a$ between 1970 and 2009 was 83 dollars per megajoule, where dollars are expressed in inflation-adjusted 2005 MER units. The mean value for the rate of return for energy consumption and wealth, $\eta = \lambda f$, over this period was 1.87 percent per year. On top of this mean was the trend in $\eta$, which increased from 1.37 percent per year in 1970 to 2.14 percent per year in 2009. So, 
the fitted innovation rate for this time period was $d\ln\eta/dt=$
0.93 \% yr$^{-1}$. This implies an average real GDP growth rate for the period of $d\ln Y/dt=$ 1.87 $+$ 0.93$=$ 2.80 \% per year. The actual observed mean was 2.93\% per year, a difference off just 0.13\% per year. Thus, accurate forecasts of global GDP growth can be inferred knowing only how fast energy productivity is improving, and without having to explicitly represent nations, sectors, people or their lifestyles.

\subsection*{Long-term forecasting of wealth}
If real innovation is positive, then from Eq. \ref{eq:eta} and Eq. \ref{eq:etaf},
the deterministic solution for the growth of wealth $C$ is
\begin{equation}
C\left(t\right)=C_{0}e^{\eta\tau_{\eta}\left(e^{t/\tau_{\eta}}-1\right)}\label{eq:X-1}\end{equation}
$C_{0}$ is todays wealth, and $\tau_{\eta}$ represents the characteristic innovation time \begin{equation}
\tau_{\eta}=\frac{1}{\rm{innovation\,rate}}=\frac{1}{d\ln\eta/dt}\label{eq:taua-1}\end{equation}
Note that the solution for $C\left(t\right)$ 
condenses to the simple exponential growth form of $C=C_{0}\exp\eta t$
in the limit that the innovation rate slows to zero and $\tau_{\eta}\rightarrow\infty$.
If there is positive innovation, however, then $\tau_{\eta}$ is positive and
finite, and wealth growth is explosive or super-exponential (i.e. the exponent of an exponent) \citet{Garrettmodes2012}. 

While more mathematically cumbersome, a familiar way of expressing growth
is in terms of doubling-times. The doubling times $\delta$ for wealth
$C$, and the rate of return $\eta$, are given respectively by
$\delta_{C}={\ln2}/{\eta}$ and $\delta_{\eta}=\tau_\eta{\ln2}$. 

Effectively $\delta_{C}$ represents the time it takes for civilization
to double its wealth, assuming current rates of return hold. Similarly, $\delta_{\eta}$ is the
time required for the rate of return to double (or $\delta_{C}$ to halve), assuming current innovation rates stay fixed.
The recent history for these quantities is shown in 
Figure \ref{fig:doubling}. From Eq. \ref{eq:X-1}, a deterministic solution for the evolution of wealth is\begin{equation}
C\left(t\right)=C_{0}2^{\frac{\delta\eta}{\delta_{C}}\left(2^{t/\delta_{\eta}}-1\right)}\label{eq:X2}\end{equation}

\newpage

\bibliographystyle{copernicus} 
\bibliography{References}

\end{document}